*Application Note*

# WinPCA: A package for windowed principal component analysis

L. Moritz Blumer[1,2,*] Jeffrey M. Good[2] and Richard Durbin[1]

[1] Department of Genetics, University of Cambridge, United Kingdom
[2] Division of Biological Sciences, University of Montana, MT, United States
[*] corresponding author: lmb215@cam.ac.uk

## Abstract

Principal component analysis (PCA) is routinely used in population genetics to assess genetic structure. With chromosomal reference genomes and population-scale whole genome-sequencing becoming increasingly accessible, contemporary studies often include characterizations of the genomic landscape as it varies along chromosomes, commonly termed genome scans. While traditional summary statistics like $F_{ST}$ and $d_{XY}$ remain integral to characterizing the genomic divergence profile, PCA fundamentally differs by providing single-sample resolution, thereby making results intuitively interpretable to help identify polymorphic inversions, introgression and other types of divergent sequence. Here, we introduce WinPCA, a user-friendly package to compute, polarize and visualize genetic principal components in windows along the genome. To accommodate low-coverage whole genome-sequencing datasets, WinPCA can optionally make use of PCAngsd methods to compute principal components in a genotype likelihood framework. WinPCA accepts variant data in either VCF or BEAGLE format and can generate rich plots for interactive data exploration and downstream presentation.

## Introduction

The combined actions of selection, gene flow, drift and recombination shape the genomic landscape of populations, causing differences in local genetic composition along chromosomes (1). Principal component analysis (PCA) has become solidly established in population genetics since initial applications to large-scale human datasets alongside the publication of the EIGENSOFT software (2) and is a cornerstone of genetic variation studies today. Inspired by previous works, especially Li & Ralph (2019) (3) and Jay et al. (2021) (4), we developed WinPCA, a software tool to compute and visualize genetic principal components from biallelic whole genome sequencing (WGS) datasets in windows along chromosomes. Versions of WinPCA have been used to detect chromosomal inversions (5) and to characterize the mosaic structure of a mouse hybrid's genome (6). WinPCA addresses two of the key challenges frequently encountered with windowed PCA, (1) the treatment of missing data and (2) the polarization of principal components along chromosomes. Since PCA algorithms inherently require complete data sets (7), variants with any missing genotype calls are typically discarded from the input. Especially with large sample datasets this can quickly result in the exclusion of most variants. WinPCA by default mean-imputes missing genotype calls like other popular PCA softwares. Alternatively, avoiding genotype imputation and at the same time accommodating low coverage WGS datasets, WinPCA can operate in a genotype likelihood framework by internally employing PCAngsd (8) machinery. WinPCA scales to large datasets – we have applied it to whole genome variant callsets of thousands of samples.

## Methods

WinPCA is written in Python and designed to be run from the UNIX command line. It is structured in a modular way, partitioning the windowed PC analysis into multiple steps from the computation of principal components to the creation of interactive output plots. Subcommands are called by typing `winpca subcommand`. The minimal input is a variant file in VCF, TSV or BEAGLE (as used by the ANGSD suite (9)) format. Below, we provide an overview of the modules.

### *winpca pca*

The core module, `pca`, parses variant files containing either hard-called genotypes or genotype likelihoods. Input files may be gzip or bgzip-compressed (10). Biallellic variants are parsed in a rolling window fashion and a principal component analysis is conducted per window using either scikit-allel (11) for hard-called genotypes or PCAngsd (8) for genotype likelihoods (GL or PL format). Particularly for low coverage WGS datasets or datasets with large numbers of samples, for which hard genotype callers such as bcftools (12) or GATK (13) frequently call some genotypes as missing at a large fraction of sites, we recommend supplying properly imputed genotype call sets or genotype likelihoods as input to WinPCA (see (14) and (15) for reference panel-free genotype imputation methods). When this is not done, WinPCA imputes missing genotype calls as the mean value on the fly. Despite limitations, this is the default for other commonly used genetic PCA algorithms (e.g. PLINK (16)), and is arguably preferable to discarding large amounts of information. Rolling windows are sequentially processed to limit memory requirements, and eigenvectors for the first two principal components are computed, alongside metadata, per window. The results are then written to a set of gzipped text files that can be used by subsequent modules.

### *winpca polarize* and *winpca flip*

Since PCA is performed separately for each genomic window, the polarity of a given PC is inhomogeneous across the output. To reduce the resulting noise for visualization purposes, WinPCA `polarize` attempts to harmonize the sign polarity of eigenvectors across adjacent windows. By default it does this by adaptively selecting the sample with the highest absolute PC value and using the sign average of the five previous (and already polarized) windows to decide whether to flip the current window. We call this adaptive auto-selection.

However, this does not always give the desired result, so we provide a variety of alternatives. First, it is possible to change the number of windows considered away from five. Alternatively, polarization can be determined by the user specifying one or more fixed guide samples. For example where a natural outgroup is contained in the sample set, it makes sense for one or a group of samples to remain consistently positive or negative in PC space. While polarization by default is performed as part of the `pca` module, output data from an existing pca run may be re-polarized by invoking the `polarize` module directly, updating the output files. The `flip` module allows to reflect PCs for an entire chromosome to homogenize their orientation across the genome and furthermore accepts a list of chromosomal coordinates to flip individual windows if incorrectly polarized regions remain after automatic polarization.

### `winpca chromplot` and `winpca genomeplot`

The plotting modules `chromplot` and `genomeplot` generate interactive plots which may be annotated with information from a user-provided metadata file. The plotly library (17) is used internally, which can generate a variety of output formats. Static plots may also be exported in vector (PDF, SVG) or raster formats (PNG), while HTML output files can be viewed interactively in a web browser. In this case metadata annotations are displayed when hovering over data points and one may zoom in and out of the plotting area or hide and view data groups. This allows for great flexibility in data exploration, while reducing the need for repeated plot generation at different scales or with alternative annotations. While `chromplot` visualizes a PC along a single chromosome, `genomeplot` jointly visualizes data across multiple chromosomes. Plotting colors, plotting order and plot interval can be controlled by the user. Quantitative metadata (e.g. phenotype scores) can be supplied to color-code samples using a continuous spectrum, which can help visualize quantitative trait loci (QTL).

## Results

We illustrate the functionality and versatility of WinPCA analyses using four publicly available datasets. These vary in taxonomic group, number of samples, sequencing depth (called genotypes vs. genotype likelihoods) and divergence level. Table 1 summarizes the key characteristics of each dataset. All analyses were conducted with the default window size of 1 Mbp if not stated otherwise.

In the first example we used WinPCA to visualize the divergence landscape across 26 human populations (2,504 individuals) that were sequenced for the 1000 Genomes Project (18) (Fig. 1A). We then focused on the well-studied 17q21 inversion region with an increased resolution (100 kb window size) local scan (Fig. 1B). The core inversion spans 900 kb, is polymorphic in many human populations with elevated frequencies in some European groups and the derived H2 haplotype has been associated with multiple disease phenotypes (19). The visible tripartite stratification into the three possible haplotype combinations is characteristic for PCA in inversion-polymorphic genomic regions (20).

We next employed WinPCA to a dataset of 110 whole genome-sequenced *Cannabis sativa* accessions representing the global diversity of wild and cultivated lineages (21). In several regions on chromosomes 1 and 4 most wild lineages diverge from hemp and drug-type strains that share a history of cultivation (Fig. 1C). These regions overlap with previously identified genomic segments of increased $F_{ST}$ (22). We noticed an additional highly divergent region on chromosome 9 and performed an increased-resolution (500 kbp window size) local scan of this region (Fig. 1D). For reference, we included a hover display item from the HTML version of this plot. The stratified segregation pattern at approx. 0 to 5 Mbp suggests the presence of a megabase-scale inversion polymorphism that segregates among all four clades but went undetected in previous $F_{ST}$ scans. Interestingly, among the included samples, only hemp-type and wild *Cannabis* representatives carry the putative homozygous inverted genotype and the region is overrepresented among positively selected genes identified in a separate dataset (Woods et al. (22), Supplementary Table 4). Some of these genes relate to traits relevant in plant environmental adaptation and cultivation (e.g., WRKY26 has a function in thermotolerance (23) and UVR8 is a photoreceptor that modulates the response to ultraviolet radiation (24)). A recent preprint reports this inversion in an alignment-based approach, thereby confirming its presence (25).

**Table 1.** Key characteristics of the analyzed datasets.

| Dataset | n | Variant type | File format | Missingness |
|---|---|---|---|---|
| Humans | 2,504 | GT | VCF | imputed |
| *Cannabis* | 110 | GT | VCF | unimputed |
| Cichlid cross | 291 | GT | VCF | unimputed |
| Mouse hybrid | 71 | GL | BEAGLE | - |

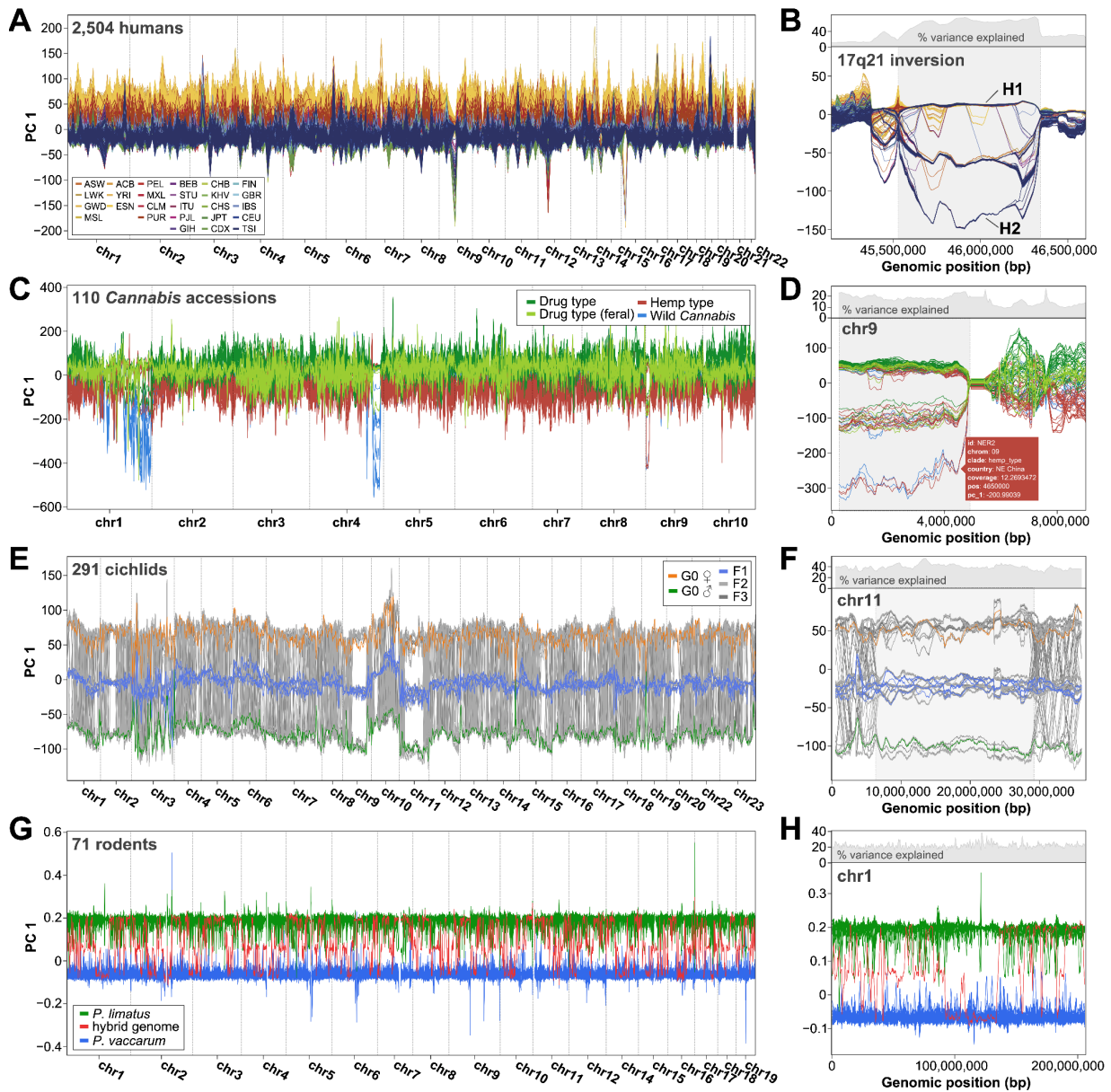

**Fig. 1: WinPCA use cases.** We applied WinPCA to four exemplary datasets. **A**, **B** 2,504 humans from the 1000 Genomes Project color-coded by population. **B** Inversion polymorphism at 17q21 (highlighted in grey) at increased resolution with H1 and H2 haplotypes annotated. **C**, **D** 110 whole genome-sequenced *Cannabis* accessions including wild *Cannabis*, hemp-type and drug-type strains. **D** Polymorphic inversion spanning approx. 0 to 5 Mbp (highlighted in grey) on chromosome 9 at increased resolution. For reference, metadata displayed when hovering over a trace in the interactive HTML version of this plot is displayed. **E**, **F** 291 cichlid genomes from an interspecific cross (up to generation F3) between *Aulonocara stuartgranti* and *Astatotilapia calliptera*. **F** Separate display of chromosome 11. While F1 individuals localize intermediate between the parents in PC space, individual recombinations can be traced in F2 and F3 individuals. Full recombination suppression is evident in the inversion region (highlighted in grey). **G**, **H** Genomes from 70 *Phyllotis vaccarum* and *P. limatus* individuals and that of a single hybrid individual highlighted in red. The mosaic ancestry of the hybrid suggests an early but advanced stage of hybridization between both species (see text).

Third, we include a dataset from our recent study of large chromosomal inversion polymorphisms in Lake Malawi haplochromine cichlids (5). Fig. 1E features 291 individuals from an interspecific cross up to generation F3 between *Aulonocara stuartgranti* and *Astatotilapia calliptera*. Individuals of the F1 generation combine one copy of each parental genome and consistently localize intermediate in PC space. In contrast, individuals from subsequent generations that underwent recombinations alternate between tracts that are homozygous for either parent's ancestry or heterozygous. Individual recombination events are distinguishable at higher resolution (Fig. 1F) and regions of recombination suppression, especially inversions on chromosomes 2, 9, 11 and 20 (5) can be easily spotted (Fig. 1E, F).

Finally, we analyzed a low coverage dataset (median: 2.58X) comprising 71 genomes of leaf-eared mice (*Phyllotis*) that is featured in our recent study (6). WinPCA was used to visualize the mosaic ancestry of an individual from a *P. vaccarum* - *P. limatus* hybrid zone (Fig. 1E, F). The hybrid genome alternates between three types of local ancestries, comparable to the previously described cichlid cross. The presence of multiple short tracts with homozygous ancestry for each parental species suggests that the hybrid individual is at an advanced stage of intercross rather than a first generation hybrid. Given the approximately equal ancestry contributions from both *P. vaccarum* and *P. limatus*, the presence of a hybrid population could be inferred, although only a single specimen from that population has been sequenced (6).

WinPCA is freely available at https://github.com/MoritzBlumer/winpca and comes with limited Python dependencies. We include a detailed step by step usage tutorial on the github repository (https://github.com/MoritzBlumer/winpca/wiki).

## Discussion

Windowed principal component analysis represents a means to examine the genomic divergence landscape without prior knowledge of population structure and allows to characterize datasets at single-sample resolution.

While for $F_{ST}$-like approaches incorrect assignment of samples to populations can mask locally increased genetic divergence (and likewise trans-population polymorphisms such as inversions may be hard to detect), windowed PCA can often overcome these downfalls. However, users should be aware of its own limitations. Principal component analysis is a dimensionality reduction method and the relevant axes of variation may be distributed across several of the higher order principal components (26). WinPCA calculates PC1 and PC2 by default, but often output plots of PC2 remain noisy despite polarization efforts and provide limited insights. We have observed that known inversion polymorphisms are usually represented in PC1, but this might not always be the case and depends on the structure of the dataset. Due to the effect of noise in genomic variant data we recommend relatively large window sizes (100,000 kbp to 1 Mbp) for genome-wide scans and emphasize the benefits of high quality reference genomes with a low degree of fragmentation and misassembly. While WinPCA can parse genotype likelihoods from VCF files, we have observed superior results when supplying BEAGLE files that were generated and filtered according to ANGSD principles (9) and we recommend this approach.

We view WinPCA as an easy-to-use, scalable tool for initial unbiased exploration of WGS datasets (low or high coverage) and as a powerful method to identify and visualize divergent genomic regions reflecting various underlying genetic causes, as illustrated in the described examples.

## Data availability

All genetic datasets used are available from public databases. 2,504 human genomes: http://ftp.1000genomes.ebi.ac.uk/vol1/ftp/data_collections/1000G_2504_high_coverage/working/20220422_3202_phased_SNV_INDEL_SV/; 110 Cannabis genomes: doi.org/10.5061/dryad.9p8cz8wfr; 291 cichlid genomes: <will be added before final publication>; 71 rodent genomes: <will be added before final publication>. Metadata used to annotate plots were derived from the associated publications and are available from https://github.com/MoritzBlumer/winpca/wiki/4--%7C--More-use-cases alongside the complete code used to generate Fig. 1 panels. For reference, interactive versions of figure panels A and D are available for download at https://github.com/MoritzBlumer/misc/raw/refs/heads/main/html_plots.zip.

## Competing interests

No competing interests are declared.

## Author contributions

L.M.B. developed the software, conducted all analyses and wrote the initial manuscript. R.D. and J.M.G. supervised the project, contributed to manuscript writing and provided funding.

## Acknowledgments

The idea to visualize principal components along chromosomes was inspired by Paul Jay et al. (4) and quickly became adopted by me and others in the group for the initial exploration of new datasets. The authors furthermore thank Jonas Meisner and Alistair Miles for extensive feedback and recommendations on how to best integrate PCAngsd and scikit-allel. Furthermore, the authors appreciate feedback and comments from Trevor Cousins, Bjørghild Breistein, Marilou Boddé, Mara Lawniczak, Vanessa Pahl and Hadi Khan. The authors gratefully acknowledge support through the Wellcome Trust (Wellcome grant 207492 to R.D) and the National Institutes of Health (R01 HL159061 to J.M.G.), and the National Science Foundation (OIA-1736249 to J.M.G.). L.M.B. was supported through a Harding Distinguished Postgraduate Scholarship and the National Science Foundation (OIA- 1736249).